\documentstyle[festkoer,epsf]{vieweg}

\makeatletter\@addtoreset{equation}{section}\makeatother

\author{D. Vollhardt, N. Bl\"umer, K. Held, M. Kollar, J. Schlipf, M. Ulmke,
and J. Wahle$^*$}

\Kurzautor{Vollhardt et al.}

\title{Metallic ferromagnetism: Progress in our understanding of an
old strong-coupling problem}

\Kurztitel{Metallic ferromagnetism}

\Adresse{Theoretische Physik III,
Elektronische Korrelationen und Magnetismus,
Universit\"at Augsburg, D-86135 Augsburg, Germany}

\begin{document}
  \Titel
  \begin{abstract}
    Metallic ferromagnetism is in general an intermediate to strong
    coupling phenomenon. Since there do not exist systematic analytic
    methods to investigate such types of problems, the microscopic
    origin of metallic ferromagnetism is still not sufficiently
    understood. However, during the last two or three years remarkable
    progress was made in this field: It is now certain that even in
    the one-band Hubbard model metallic ferromagnetism is stable in
    dimensions $d=1,$ $2$, and $\infty$ on regular lattices and at
    intermediate values of the interaction $U$ and density $n$. In
    this paper the basic questions and recent insights regarding the
    microscopic conditions favoring metallic ferromagnetism in this
    model are reviewed. These findings are contrasted with the results
    for the orbitally degenerate case.
  \end{abstract}

  \section{Introduction}
  What is the microscopic origin of ferromagnetism?  Exactly seventy
  years ago, in 1928, Heisenberg addressed this
  question~\cite{Heisenberg28} after having discovered the phenomenon
  of quantum mechanical exchange and the corresponding exchange
  interaction. He formulated a spin model (the Heisenberg model),
  hoping to be able to answer precisely this question. However, it was
  pointed out by Bloch~\cite{Bloch29} that a model of localized spins
  cannot explain {\it metallic} ferromagnetism as observed in iron,
  cobalt, and nickel, and that a proper model would have to include
  the itineracy of the electrons, i.e. the band aspect. Based on the
  observation that the Curie temperature $T_c \sim 10^3\mbox{~K} \sim
  0.1$~eV in these systems it is clear that the kinetic
 energy and  the spin-{\it in}dependent Coulomb
  interaction, together with the Pauli principle, must ultimately be
  responsible for metallic ferromagnetism. Ever since one has been
  looking for the simplest microscopic model and mechanism explaining
  the origin of metallic ferromagnetism and, equally important, for
  analytic solutions or at least controlled approximations for these
  models~\cite{Mattis88}. Today we know that even with the ``right"
  model these answers are not easily obtained since metallic
  ferromagnetism generally occurs only at {\it intermediate} to {\it
    strong} coupling and off half filling~\cite{LiebFaz,Vollhardt97b}.
  Thus, it belongs to the class of problems for which systematic
  theoretical approaches do not exist. Namely, weak-coupling theories
  or renormalization group approaches~\cite{Shankar94} 
  which are so effective in
  detecting instabilities with respect to antiferromagnetism or
  superconductivity, do not work in this case.  Instead, {\it
    nonperturbative} methods are required.
  
  During the last two or three years significant progress was made in
  our understanding of the microscopic foundations of metallic
  ferromagnetism.  These insights were made possible by both new
  analytic methods and new numerical techniques. In this paper some of
  these recent developments will be reviewed.  In particular, the
  microscopic conditions for metallic ferromagnetism in the one-band
  Hubbard model (Section \ref{sec:oneband}) and in the case of orbital
  degeneracy (Section \ref{sec:zweiband}) are explained and the
  differences discussed.  A conclusion (Section \ref{sec:concl})
  closes the presentation.

  \section{The one-band Hubbard model}\label{sec:oneband}
\begin{figure}
    \epsfxsize=0.8\hsize \hspace{0.1\hsize}\epsfbox{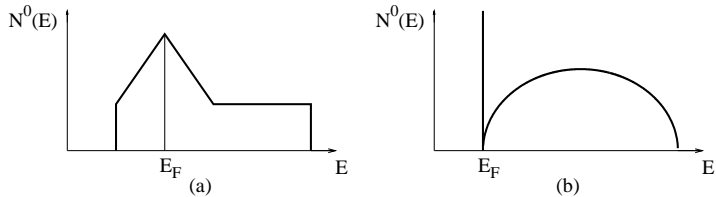}
    \caption{(a) Model DOS favoring ferromagnetism in the Hubbard model as
      suggested by Kanamori~\cite{Kanamori63}; (b) DOS for
      flat-band ferromagnetism (schematic)~\cite{Mielke93b}.
	\label{fig:kirchen}}
  \end{figure}

  The simplest lattice model for correlated electrons, the one-band
  Hubbard model
  \begin{equation}\label{eqn:hub}
    {H}_{\mbox{\scriptsize Hub}} = -\sum_{i,j,\sigma} t_{ij}({c
      }^\dagger_{i\sigma} {c}^{\phantom \dagger}_{j\sigma}+ {\rm
      h.c.})+ U \sum_{i} {n}_{i\uparrow} {n}_{i\downarrow} \label{hub}
  \end{equation}
  was proposed independently by Gutzwiller~\cite{Gutzwiller63},
  Hubbard~\cite{Hubbard63}, and Kanamori~\cite{Kanamori63} in 1963,
  with the explanation of metallic ferromagnetism in 3d transition
  metals in mind. Concerning the suitability of (\ref{eqn:hub}) to
  describe metallic ferromagnetism for general $U$ and electron
  densities $n$ the three authors came to different conclusions. In
  any case, the theoretical methods used at that time were not
  controlled enough to provide definitive conclusions. This is also
  true for most of the research following their original work, with a
  few exceptions such as Nagaoka's theorem for a single hole at
  $U=\infty$~\cite{Nagaoka66}. We note that in the past the kinetic
  energy in (\ref{eqn:hub}) was usually restricted to nearest-neighbor
  (NN) hopping; then it is useful to divide the underlying lattices
  into bipartite and nonbipartite ones. About ten years ago the
  interest in the subject started to rise again~\cite{Lieb89}. In
  particular, by reducing Kanamori's~\cite{Kanamori63} model density
  of states (DOS) of noninteracting electrons, $N^0(E)$
  (Fig.~\ref{fig:kirchen}a), to its barest minimum
(Fig.~\ref{fig:kirchen}b) Mielke~\cite{Mielke91} began to investigate
  the stability of ferromagnetism in systems with flat, i.e.
  dispersionless, bands.  He~\cite{Miel} and Tasaki~\cite{Tasaki92}
  were able to derive rigorous criteria for the existence of
  ferromagnetism in these particular systems~\cite{Mielke93b}.  
  Generalizations to
  nearly-flat bands are also possible~\cite{Tas}.  Ferromagnetism is
  proven to exist when the lowest band is half-filled and the system
  is insulating, as well as close to half filling.  Due to the
  pathological degeneracy of the ground state it is still not exactly
  clear whether away from half filling one really obtains {\it
    metallic} ferromagnetism (for a detailed discussion see
  Ref.~\cite{Tasaki98}).
             
  A different route to ferromagnetism was taken by
  M\"uller-Hartmann~\cite{MuellerHartmann95} who investigated the
  $t$-$t'$ Hubbard model
  \begin{equation}
    {H}_{\mbox{\scriptsize Hub}}^{t t'} = 
    - t \sum_{{\mbox{\scriptsize NN,}}\sigma} 
    ({c}^\dagger_{i\sigma} {c}^{\phantom \dagger}_{j\sigma}+ {\rm h.c.})
    - t' \sum_{{\mbox{\scriptsize NNN,}}\sigma} 
    ({c}^\dagger_{i\sigma} {c}^{\phantom \dagger}_{j\sigma}+ {\rm h.c.})
    + U \sum_{i} {n}_{i\uparrow} {n}_{i\downarrow},
  \end{equation} 
  i.e.~model (\ref{eqn:hub}) with NN and next-nearest neighbor (NNN)
  hopping.  He found that in $d=1$ at $U=\infty$ ferromagnetism
  becomes possible in the low-density limit ($n \rightarrow 0$). This
  scenario was extended by Pieri et al.~\cite{Pieri96} and, in
  particular, by Penc et al.~\cite{Penc96} who introduced a
  generalized model which can be shown to have a metallic phase in
  $d=1$.
  
  In the Hubbard model the interaction term is completely independent
of lattice and
  dimension. Therefore the kinetic energy, or dispersion,
  of the electrons and the underlying lattice must play an important
  role for the stability of metallic ferromagnetism. This is indeed
  seen explicitly in all of the above-mentioned investigations and is
  also apparent in the studies of the single spin-flip instability of
  the Nagaoka state for which Hanisch et al.~\cite{Hanisch97} recently
  derived significantly improved bounds for various lattices in $d=2$
  in $d=3$, and which was solved analytically by Uhrig~\cite{Uhrig96}
  in the limit of $d=\infty$ for several nonbipartite lattices.

  \subsection{Routes to ferromagnetism}
\begin{figure}
    \epsfxsize=0.6\hsize\hspace{0.2\hsize}\epsfbox{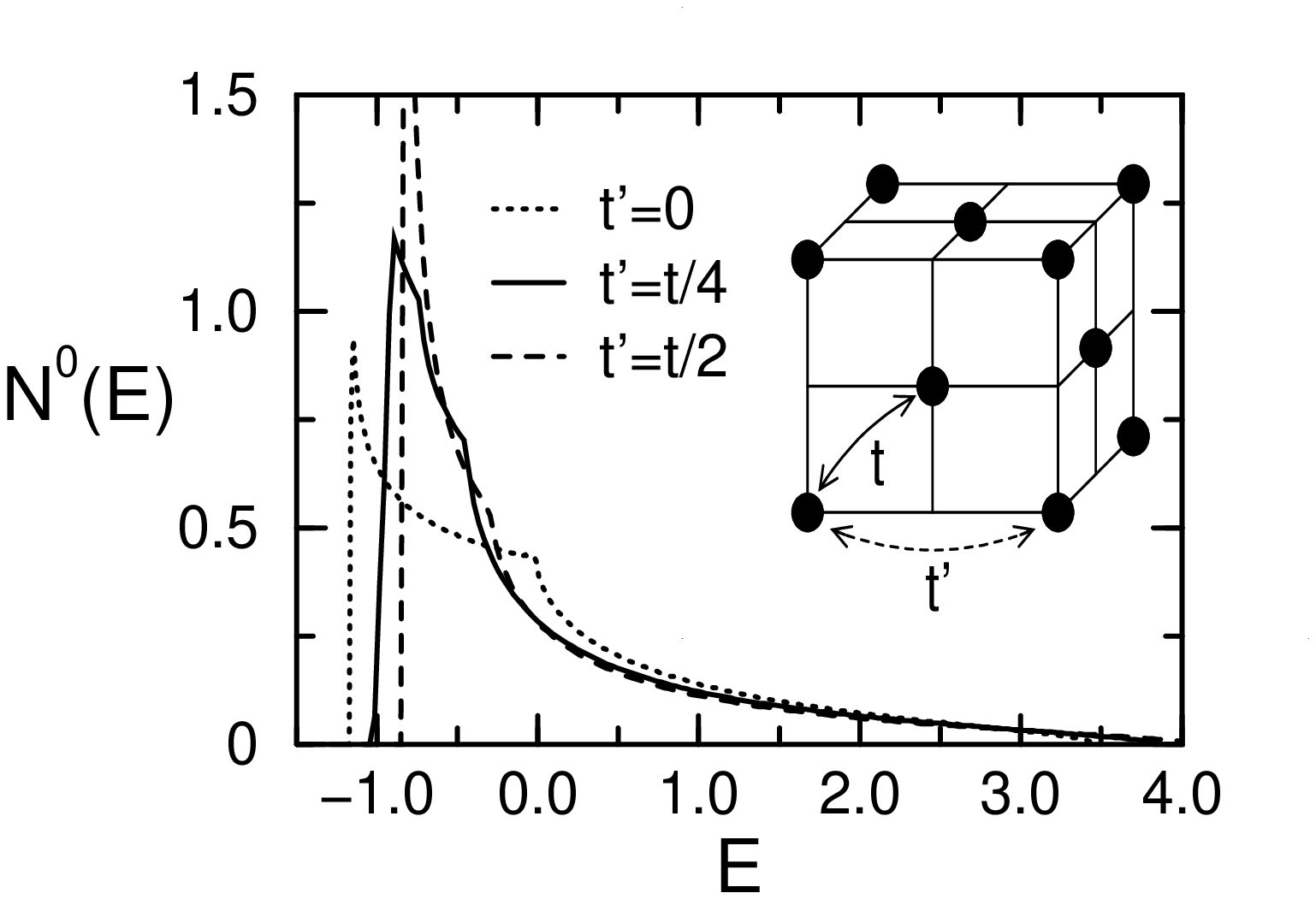}
    \caption{DOS of noninteracting electrons on a fcc lattice in $d=3$
      with and without additional NNN-hopping $t'$.\label{fig:fccdos}}
  \end{figure}

  On bipartite lattices the $t'$-hopping term destroys the
  antiferromagnetic nesting instability at small
  $U$~\cite{Hofstetter98}. In $d > 1$ it shifts spectral weight to the
  band edges and thereby introduces an asymmetry into the otherwise
  symmetric DOS. It will be shown below that a high spectral weight at
  the band edge (more precisely: the {\it lower} band edge for $n<1$)
  minimizes the loss of kinetic energy of the overturned spins in the
  magnetic state and is hence energetically favorable.  Therefore
  frustrated, i.e.~nonbipartite lattices, or bipartite lattices with
  frustration due to hopping (e.g.~$t' \not =0$) support the
  stabilization of metallic ferromagnetism. The fcc lattice is an
  example for a frustrated lattice in $d=3$. The corresponding DOS of
  the noninteracting particles is shown in Fig.~\ref{fig:fccdos}.
  Switching on an additional NNN hopping $t'$ is seen to further
  increase the spectral weight at the lower band edge. For $t'= t/2$
  one even obtains a square-root-like divergence.
  
\begin{figure}
    \epsfxsize=\hsize\epsfbox{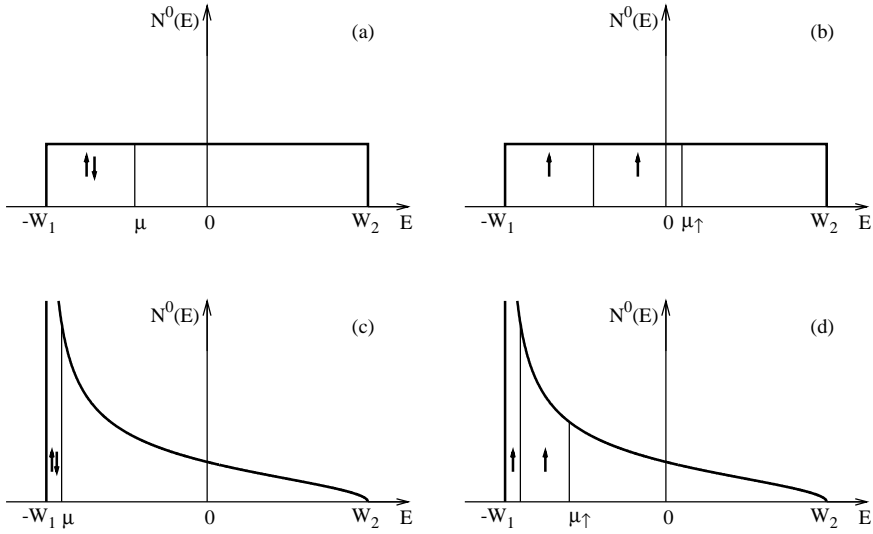}
    \caption{Flat, symmetric DOS for (a) unpolarized and (b) fully
      polarized electrons; (c) and (d): same as in (a), (b) but for a
      strongly peaked DOS.\label{fig:kinerg}}
  \end{figure}

  To understand why a high spectral weight at the band edge is
  favorable for the kinetic energy we first consider the case $U=0$,
  $n<1$~\cite{Vollhardt97b}. Let us consider a flat, symmetric DOS as
  in Figs.~\ref{fig:kinerg}a, b. Fig.~\ref{fig:kinerg}a describes the
  paramagnetic state. The fully polarized state is obtained by
  inverting the spin of the down electrons, which due to the Pauli
  principle have to occupy higher energy states. Counting the energy
  from the lower band edge the Fermi energy of the polarized state,
  $\mu_\uparrow$, is seen to be twice that of the unpolarized state
  (Fig.~\ref{fig:kinerg}b). This should be contrasted with the DOS
  having large spectral weight at the lower band edge shown in
  Figs.~\ref{fig:kinerg}c and \ref{fig:kinerg}d. Here the Fermi level
  of the polarized state is not so strongly shifted upwards,
  i.e.~fewer high energy states are populated, which is clearly
  energetically favorable. The energy difference between the fully polarized
  state and the unpolarized state
  \begin{equation}
    \Delta E = \bigg[\int\limits_{-W_1}^{\mu_\uparrow} 
    - \;2 \int\limits_{-W_1}^{\mu} \bigg]
    \;dE \;N^0(E)\; E
  \end{equation}
  must become negative for the ferromagnetic state to be stable. Of
  course, in the noninteracting case $\Delta E > 0$~\cite{Lieb62}.
  Nevertheless, even for $U=0$, $\Delta E$ attains its lowest value for a
  DOS with peaked spectral weight at the lower band edge for all
  $n$~\cite{Vollhardt97b}. To show that $\Delta E < 0$ for $U > 0$
  requires a good estimate of the energy of the {\it correlated
    paramagnet} -- this is indeed a central problem of any correlation
  theory. It should be noted that the above discussion concerning the
  shape of the DOS goes beyond the well-known Stoner criterion which
  predicts an instability of the paramagnet for $U$ equal to the
  inverse of the DOS precisely {\it at} the Fermi level.
  
  Another possibility to stabilize ferromagnetism is to consider those
  interactions which are neglected in the Hubbard interaction, in
  particular the NN direct-exchange interaction. The effect of this
  and other terms will be discussed in Section \ref{sec:nninteractions}.

  \subsection{Numerical investigation of the Hubbard model on frustrated 
    lattices in $d=1, 2$, and $\infty$} Since metallic ferromagnetism
  is an intermediate coupling problem purely analytic approaches meet
  only with limited success, in particular in dimensions $ d > 1$. In
  this situation the development of new numerical techniques in the
  last few years was of crucial importance for progress in this field.
  In particular, the density matrix renormalization group (DMRG),
  applicable mostly in $d=1$, the projector quantum Monte Carlo
  method, and the dynamic mean-field theory (DMFT), i.e. the large $d$
  limit, in connection with quantum Monte Carlo (QMC) have led to
  explicit, reliable results in dimensions $d=1, 2, \infty$.
  
\begin{figure}
    \leavevmode
    \epsfxsize=0.45\hsize\epsfbox{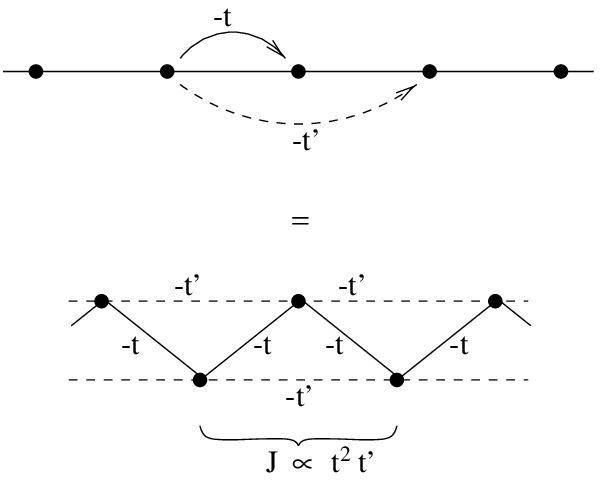}%
    \hspace{0.05\hsize}%
    \epsfxsize=0.5\hsize\epsfbox{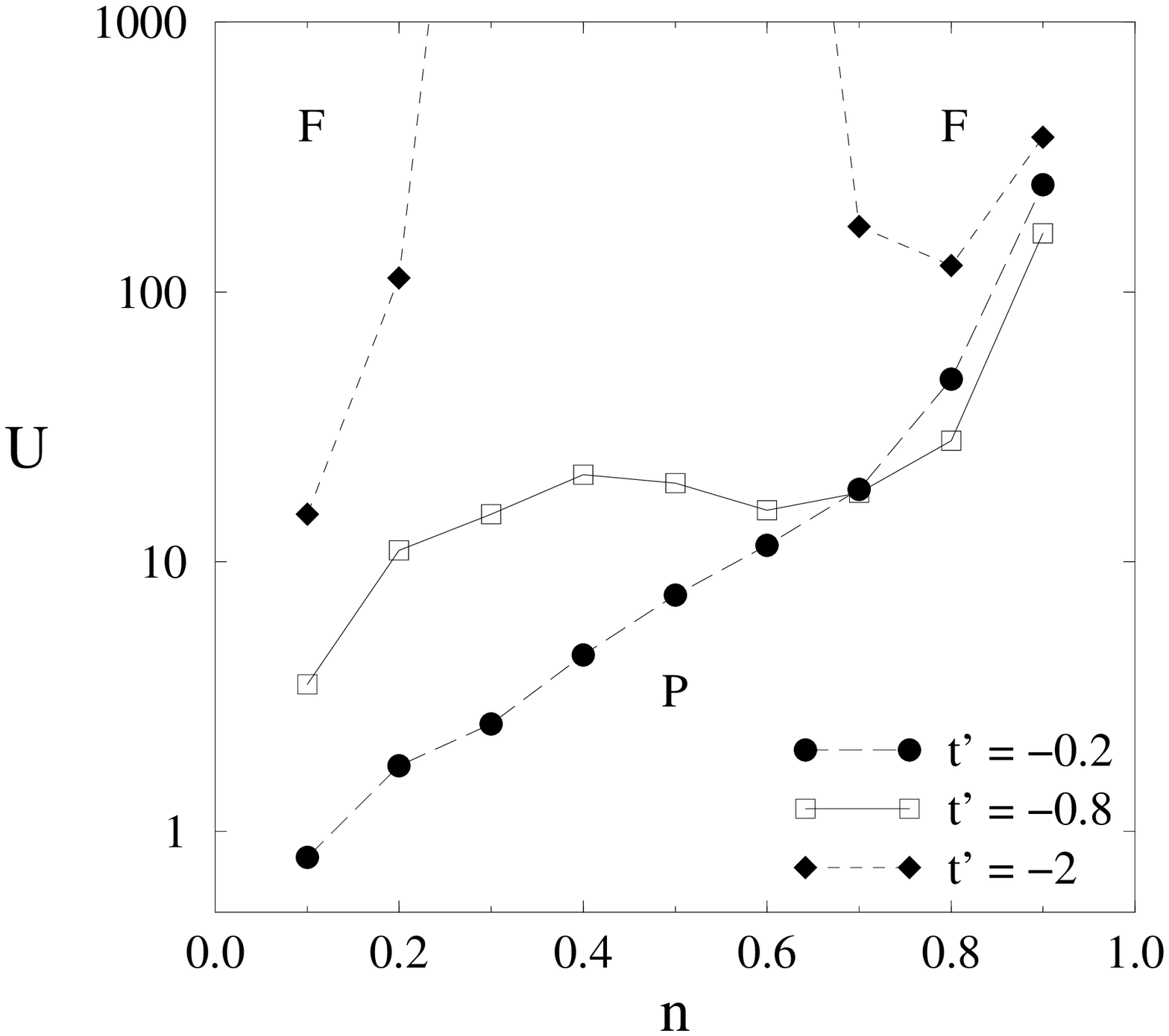}
    \par\makebox[0.45\hsize]{\small (a)}\hspace{\fill}%
    \makebox[0.55\hsize]{\small (b)}
    \caption{(a) $t$-$t'$ Hubbard model in $d=1$. (b) Ground state phase
      diagram $U$ vs.~$n$ for the $t$-$t'$ Hubbard model in $d=1$ for
      several values of $t'$ (after Daul and Noack~\cite{Daul}). P:
      paramagnetic phase, F: ferromagnetic phase.\label{fig:zigzag-dmrg}}
  \end{figure}

  ${\bf d=1}$: In one dimension the $t$-$t'$ Hubbard model may be
  viewed as a zig-zag chain made of triangular units
  (Fig.~\ref{fig:zigzag-dmrg}a).  Taking one of these units by itself
  the effective exchange interaction between the spins of two
  electrons due to hopping of an electron or hole along the triangle
  is $J \propto t^2 t'$~\cite{Penc96,Tasaki95b}.  It clearly shows
  that the sign of $t'$ is crucial: only for $t' < 0$ does one obtain
  a ferromagnetic exchange; this seems to hold even in the extended
  system. Of course, $d=1$ is a special dimension, since (i)
  ferromagnetic order is only possible at $T=0$ and $t'
  \not=0$~\cite{Lieb62}, and (ii) the DOS is always large at the band
  edges.
  
  The $t$-$t'$ Hubbard model was studied in detail by Daul and
  Noack~\cite{Daul} using DMRG for systems as large as 140 sites. The
  resulting magnetic phase diagram is shown in
  Fig.~\ref{fig:zigzag-dmrg}b. Even at small $\vert t' \vert$ a large
  ferromagnetic region in the $U$ vs.  $n$ phase diagram is found to
  exist. As $ \vert t' \vert$ increases the region of stability
  shrinks. This is due to the fact that for $ \vert t' \vert
  \rightarrow \infty$ the zig-zag chain separates into two unconnected
  chains with $t'$ as pure NN hopping; in this case the Lieb-Mattis
  theorem~\cite{Lieb62} rules out a ferromagnetic state.
  
  ${\bf d=2}$: The $t$-$t'$ Hubbard model on a square lattice was
  investigated by Hlubina et al.~\cite{Hlubina97}. At $T=0$, using
  projector QMC on systems as large as 20 x 20 sites and working at
  specific ``van Hove densities" for which the Fermi energy coincides
  with the divergence in the noninteracting DOS, these authors found a
  region of metallic ferromagnetism, e.g. at $\vert t'\vert= 0.47t$.
  
  ${\bf d=\infty}$: Already in three dimensions the coordination number
  of a fcc lattice is $Z=12$.  It is therefore quite natural to view
  $Z$ as a large number, and to consider the limit $Z \rightarrow
  \infty$~\cite{Metzner89a}. In this case one has to scale the
  hopping, e.g. as $t=t^*/\sqrt{Z}$ (in the following $t^*=1$), and
  thus obtains a purely local theory where the self energy becomes
  ${\bf k}$ independent and where the propagator $G({\bf k}, \omega) =
  G^0 ({\bf k}, \omega - \Sigma (\omega))$ may be represented by the
  noninteracting propagator at a shifted frequency~\cite{Vollhardt93}.
  In this limit the dynamics of the quantum mechanical correlation
  problem is fully included, but due to the local nature of the theory
  there is no {\it short}-range order in position space. The
  dependence on the lattice or the dispersion is then encoded in the
  DOS $N^0(E)$ of the noninteracting particles. In view of these
  properties the $d \rightarrow \infty$ limit is now generally
  referred to as ``dynamical mean-field theory''
  (DMFT)~\cite{PruschGeo}.
  
\begin{figure}
    \epsfxsize=0.5\hsize
\hspace{0.25\hsize}\epsfbox{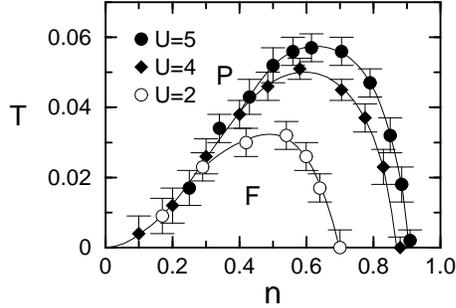}
    \caption{$T$ vs.~$n$ phase diagram of the Hubbard model for a
      fcc lattice in $d=\infty$ for several values of
      $U$~\cite{Ulmke98}.\label{fig:fccphas}}
  \end{figure}

\begin{figure}
    \leavevmode
    \epsfxsize=0.45\hsize\epsfbox{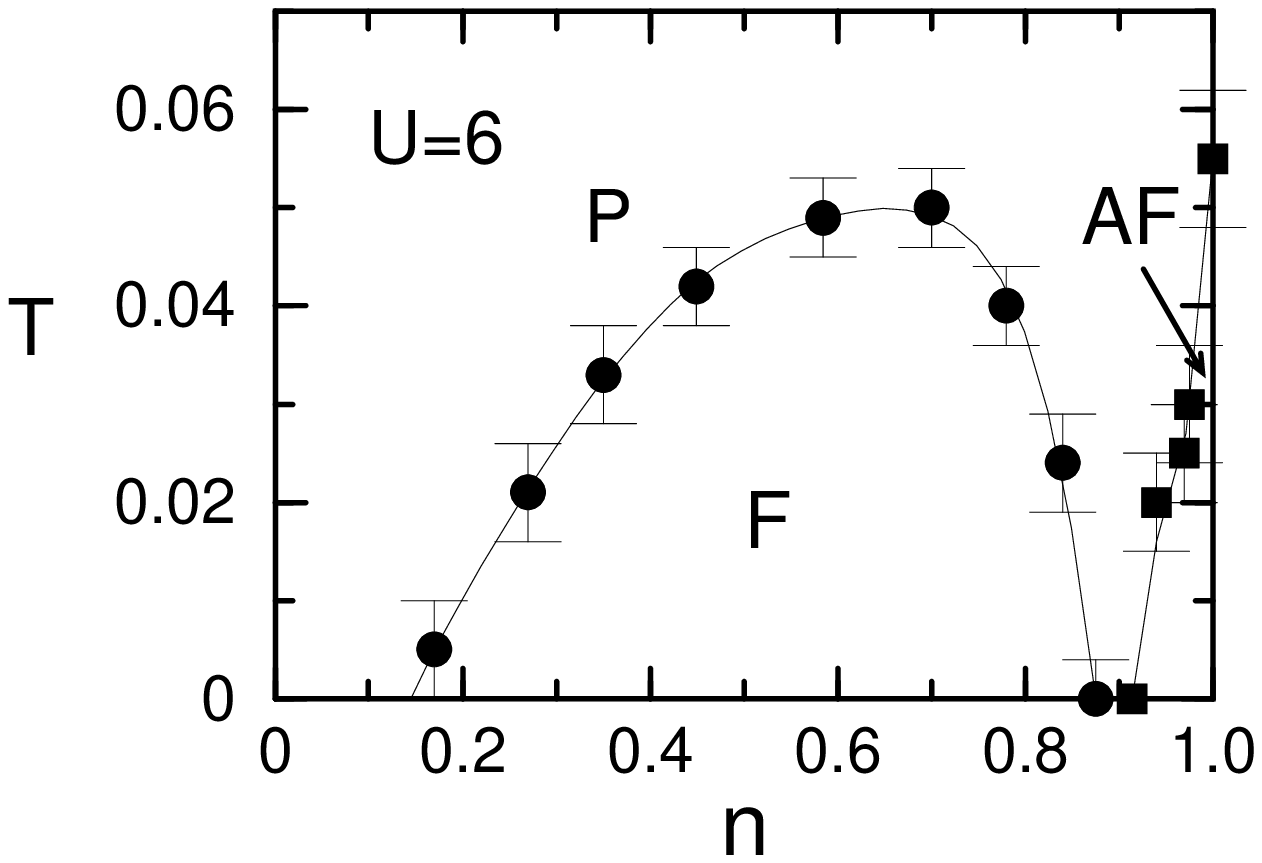}%
    \hspace{0.1\hsize}\epsfxsize=0.45\hsize\epsfbox{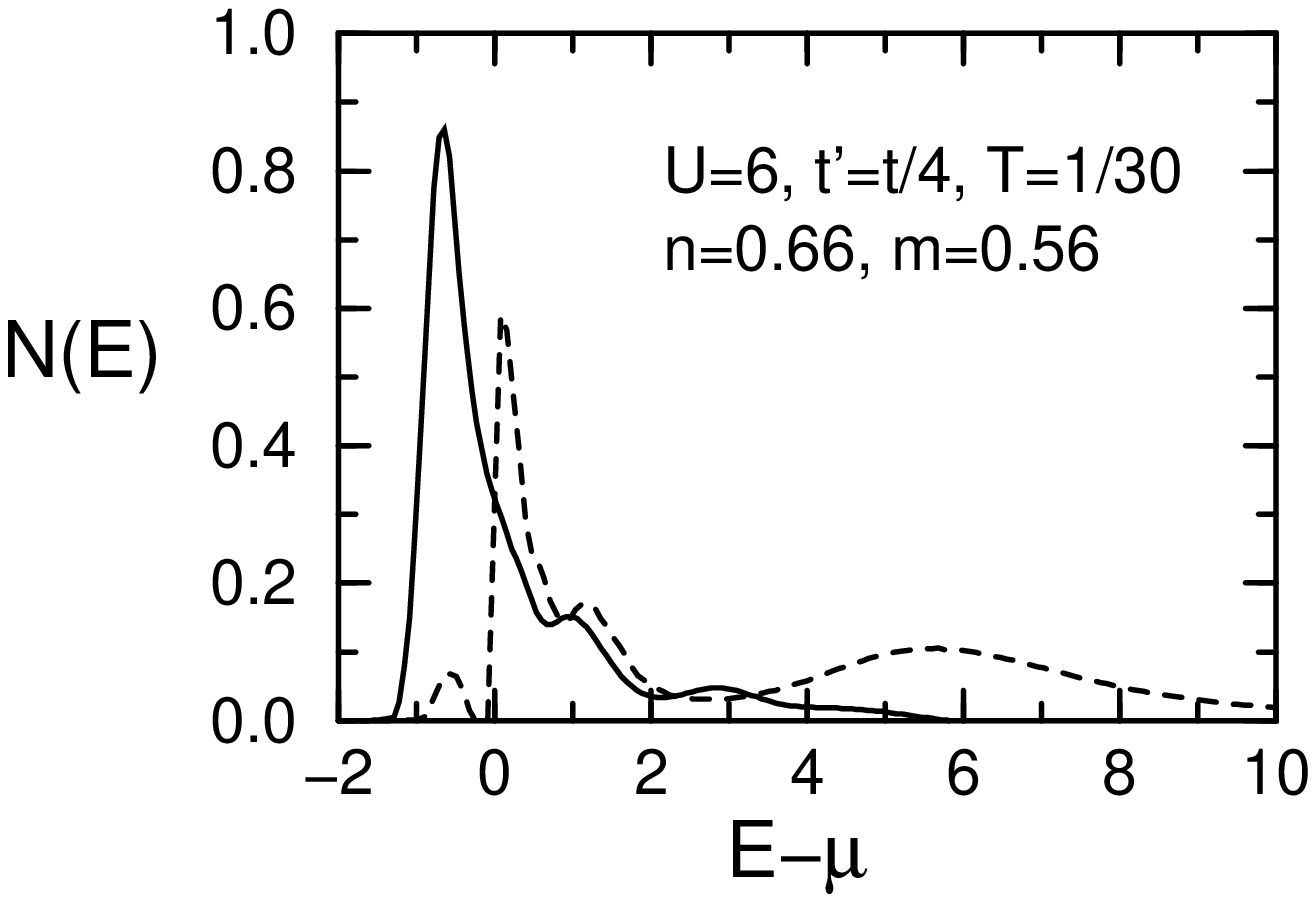}
    \par\vspace{-1.5ex}\makebox[0.45\hsize]{\small (a)}%
    \hspace{0.1\hsize}%
    \makebox[0.45\hsize]{\small (b)}
    \caption{(a) $T$ vs.~$n$ phase diagram of the Hubbard model as
      obtained within DMFT for the DOS corresponding to a
      three-dimensional fcc lattice with NN-hopping $t'=t/4$ (see
      Fig.~\ref{fig:fccdos})~\cite{Ulmke98}; 
      (b) DOS of the interacting electrons in the
ferromagnetic phase of (a), solid line: majority spin, dashed line:
minority spin.\label{fig:fcc3dim}}
  \end{figure}

  Investigations of the stability of metallic ferromagnetism on
  fcc-type lattices in large dimensions, obtained by solving the DMFT
  equations by finite-temperature QMC techniques, were first performed
  by Ulmke~\cite{Ulmke98}.  The resulting $T$ vs.~$n$ phase diagram is
  shown in Fig.~\ref{fig:fccphas} for different values of the
  interaction parameter $U$.  At $T=0$ the critical interaction
  $U_c(n)$ (see Fig.~\ref{fig:fccphas}) 
  is consistent with the analytically obtained spin-flip
  results by Uhrig~\cite{Uhrig96}. The region of stability is seen to
  increase with $U$.  By using an improved iterated perturbation
  theory to solve the DMFT equations Nolting et al.~\cite{HerrVeg}
  obtained a similar phase diagram. To make contact with $d=3$ we now
  use the
  corresponding fcc DOS shown in Fig.~\ref{fig:fccdos}.  For
  $t'=0$ no instability is found at temperatures accessible to QMC.
  However, already a small contribution of $t'$-hopping (which is
  present in any real system) is enough to produce a large region of
  stability for metallic ferromagnetism in addition to an
  antiferromagnetic phase close to half filling
  (Fig.~\ref{fig:fcc3dim}a)~\cite{Ulmke98}. This shows the strong and
  subtle dependence of the stability on the dispersion and the
  distribution of spectral weight in the DOS. The maximal transition
  temperature is $T_c^{\mbox{\scriptsize max}}= 0.05 \simeq 500$~K for
  a band width $W$ = 4 eV. This is
  well within the range of real transition temperatures, e.g. in
  nickel.

  So far we only argued on the basis of the shape of the DOS of the
  {\it non}\-in\-ter\-act\-ing electrons, $N^0(E)$. On the other hand
  the interaction will renormalize the band and relocate spectral
  weight.  Therefore it is not {\it a priori} clear at all whether the
  arguments concerning the kinetic energy etc.~(see
  Fig.~\ref{fig:kinerg}) still hold even at finite $U$. To settle this
  point we calculate the DOS of the interacting system, $N(E)$, by the
  maximum entropy method. In Fig.~\ref{fig:fcc3dim}b we show $N(E)$
  corresponding to the parameter values leading to the phase diagram
  in Fig.~\ref{fig:fcc3dim}a. Clearly the ferromagnetic system is
  metallic since there is appreciable weight at the Fermi level ($E=
  \mu$).  Furthermore, the spectrum of the majority spins is seen to
  be only slightly affected by the interaction, the overall shape of
  the noninteracting DOS being almost unchanged (the magnetization is
  quite large ($m=0.56$ at $n=0.66$) and hence the
  electrons in the majority band are almost noninteracting). This implies that
  the arguments concerning the distribution of spectral weight in the
  noninteracting case and the corresponding kinetic energy are even
  applicable to the polarized, interacting case. The spectrum of the
  minority spins is slightly shifted to higher energies and has a
  pronounced peak around $E- \mu \simeq U=6$.

  
\begin{figure}
    \epsfxsize=0.6\hsize\hspace{0.2\hsize}\epsfbox{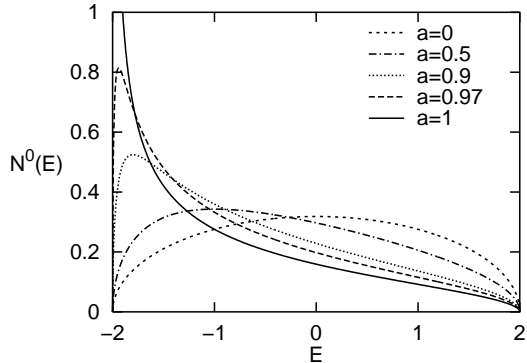}
    \caption{Model DOS, Eq.~(\ref{eqn:dos}), shown for different values of the 
      asymmetry parameter $a$~\cite{Wahle98}.\label{fig:ados}}
  \end{figure}

  To study the influence of the distribution of spectral weight on the
  stability of ferromagnetism within the DMFT systematically Wahle et
  al.~\cite{Wahle98} recently solved the DMFT equations with a tunable
  model DOS (Fig.~\ref{fig:ados}),
  \begin{equation}\label{eqn:dos}
    N^0(E)=c\,\frac{\sqrt{D^2-E^2}}{D+aE}.
  \end{equation}
  with $c=(1+\sqrt{1-a^2})/(\pi D)$ and half-bandwidth $D\equiv2$.
  Here $a$ is an asymmetry parameter which can be used to change the
  DOS continuously from a symmetric, Bethe lattice DOS ($a=0$) to a
  DOS with a square-root divergence at the lower band edge ($a=1$),
  corresponding to a fcc lattice with $t'=t/4$ in $d=3$
  (Fig.~\ref{fig:fccdos}). It is possible, in principle, to map any
  $N^0(E)$ to a dispersion $E({\bf k})$ (although not uniquely).
  
\begin{figure}
    \leavevmode
    \epsfxsize=0.57\hsize\epsfbox{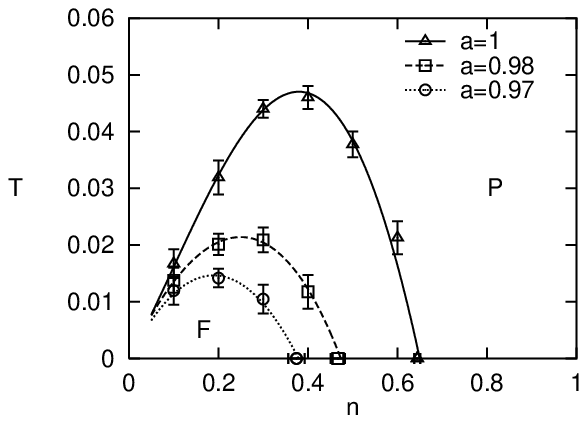}%
    \epsfxsize=0.43\hsize\epsfbox{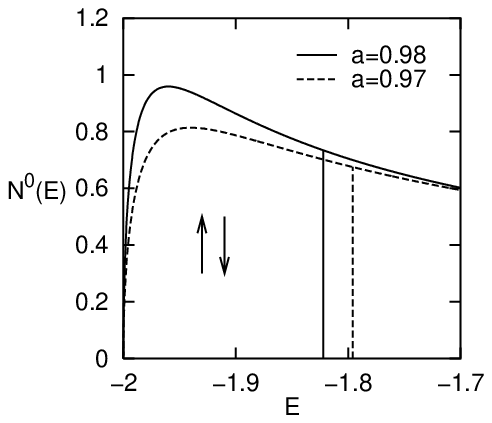}
    \par\vspace{-1.5ex}\makebox[0.6\hsize]{\small (a)}\makebox[0.4\hsize]{\small (b)}
    \caption{(a) $T$ vs.~$n$ phase diagram of the Hubbard model as
      obtained within DMFT; (b) corresponding shapes of the
noninteracting DOS; Fermi energies for $n=0.3$ are indicated by
vertical lines~\cite{Wahle98}.\label{fig:adosdmft}}
  \end{figure}

  The strong dependence of the stability of metallic ferromagnetism on
  the distribution of spectral weight is shown in
  Fig.~\ref{fig:adosdmft}a. Already a minute increase in spectral
  weight near the band edge of the noninteracting DOS, obtained by
  changing $a$ from $0.97$ to $0.98$ (see Fig.~\ref{fig:adosdmft}b) is
  enough to almost double the stability region of the ferromagnetic
  phase. It should be mentioned that Obermeier et
  al.~\cite{Obermeier97} found ferromagnetism even on a hypercubic,
  i.e.~bipartite, lattice, but only at very large $U$ values ($U>30$).
  
\begin{figure}
    \epsfxsize=0.6\hsize\hspace{0.2\hsize}\epsfbox{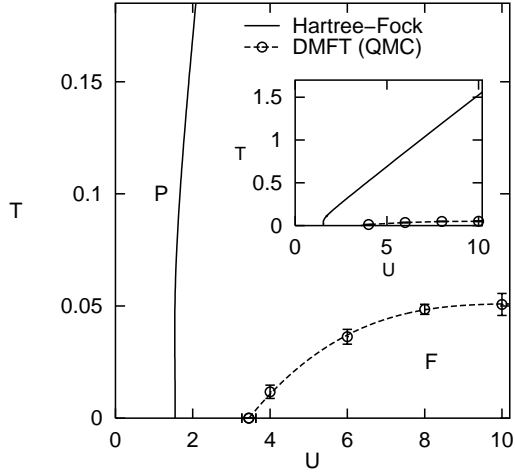}
    \caption{$T$ vs.~$U$ phase diagram for a strongly peaked DOS
      ($a=0.98$, see Fig.~\ref{fig:adosdmft}b) at $n=0.4$ within DMFT
      (circles; dashed line is guide to the eyes only) in comparison with
      Hartree-Fock (solid line)~\cite{Wahle98}.\label{fig:adostvsu}}
  \end{figure}

  The importance of genuine correlations for the stability of
  ferromagnetism is apparent from Fig.~\ref{fig:adostvsu}, where the
  DMFT results are compared with Hartree-Fock theory~\cite{Wahle98}.
  The quantum fluctuations, absent in Hartree-Fock theory, are seen to
  reduce the stability regime of ferromagnetism drastically. Spatial
  fluctuations (e.g. spin waves), absent also in the DMFT, should be
  expected to reduce that stability regime further.
  
  \subsection{Additional interactions}\label{sec:nninteractions}
  In the one-band Hubbard model only the local interaction is
  retained.  Thereby, several interactions which naturally arise when
  the Coulomb interaction is expressed in Wannier representation are
  neglected. Even in the limit of a single band and taking into
  account only NN contributions, four additional interactions
  appear~\cite{GammCamp,Hirsch,Vollhardt97b}:
  \begin{eqnarray}
    {V}^{\mbox{\scriptsize NN}}_{\mbox{\scriptsize 1-band}} & = & 
    \sum_{\mbox{\scriptsize NN}} \bigg[    
    V {n}_i  {n}_j
    + X \sum_{\sigma} ({c}^\dagger_{i\sigma} 
    {c}^{\phantom \dagger}_{j\sigma} + {\rm h.c.}) 
    ( {n}_{i,-\sigma}+ {n}_{j,-\sigma}) \nonumber\\
    & &  -2 F ({\bf {{S}}}_{i}\cdot {\bf {{S}}}_{j} 
    +\frac {1}{4} {n}_{i} {n}_{j}) \label{vnn}
    + F{'}  ({c}^\dagger_{i\uparrow} {c}^\dagger_{i\downarrow}
    {c}^{\phantom \dagger}_{j\downarrow} 
    {c}^{\phantom \dagger}_{j\uparrow}+ {\rm h.c.})\bigg].
  \end{eqnarray}
  Here the first term corresponds to a density-density interaction,
  the second term to a density-dependent hopping, and the fourth term
  describes the hopping of doubly occupied sites. In particular, the
  third term (with $F=F^*/Z>0$)
  \begin{figure}
    \epsfxsize=0.6\hsize\hspace{0.2\hsize}\epsfbox{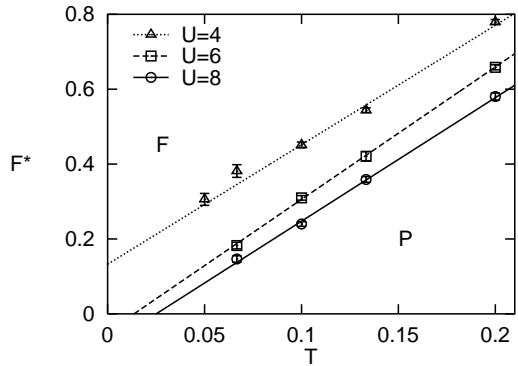}
    \caption{Direct exchange $F^*$ vs.~$T$ phase diagram for the
      generalized model $H=H_{\mbox{\scriptsize Hub}} + H_F$ for different
      values of $U$ in the case of a strongly peaked DOS ($a=0.98$, see
      Fig.~\ref{fig:adosdmft}b) at $n=0.6$. The linear extrapolation to
      $T=0$ shows that there exists a critical value of $U$ above which
      ferromagnetism is stable even in the absence of the direct
      exchange~\cite{Wahle98}.\label{fig:adosfvst}}
  \end{figure}
  \begin{equation}
    H_F = - 2 \frac{F^*}{Z} \sum_{\mbox{\scriptsize NN}} 
    {\bf {{S}}}_{i}\cdot {\bf {{S}}}_{j}
  \end{equation}
  describes the direct ferromagnetic exchange between
  electrons on NN sites. It is this spin-type interaction which
  Heisenberg in his original model singled out as the main source of
  ferromagnetism. It should be noted, however, that this interaction
  is present even when the electrons are not localized but are free to
  move.  The exchange interaction will be quite small, but
  nevertheless it favors ferromagnetic ordering in the most obvious
  way.  Hirsch~\cite{Hirsch} argued that this term is the main driving
  force for metallic ferromagnetism in systems like iron, cobalt, and
  nickel. Indeed, one can show rigorously that a next-neighbor direct
  exchange interaction, if chosen large enough, can easily trigger the
  ferromagnetic instability~\cite{StraKoll,Vollhardt97b}. To
  investigate the importance of the direct exchange interaction we
  supplement the Hubbard model by this term, $H=H_{\mbox{\scriptsize
      Hub}} + H_F$,
  neglecting all the other NN interactions. We note
  that within the DMFT the Heisenberg exchange reduces to the
  Weiss/Hartree-Fock contribution. In Fig.~\ref{fig:adosfvst} the
  influence of the exchange interaction on the stability regime of
  ferromagnetism is depicted~\cite{Wahle98}. For $F^* =0$
  ferromagnetism is unstable down to the lowest temperatures for
  $U=4$. However, by taking into account a small value of $F^*\simeq
  0.15 \ll U$ at $T=0$, the ferromagnetic phase is stabilized.  Likewise, at
  larger values of $U$ the critical temperature for the onset of
  ferromagnetism is significantly enhanced.  Hence, $F^*$ (and also the
  other neglected interactions) may well be important for
  systems on the verge of a ferromagnetic instability.  Nevertheless,
  since we now know that the Hubbard interaction $U$ together with a
  suitable kinetic energy is sufficient to trigger a ferromagnetic
  instability the ferromagnetic exchange interaction does not, in
  general, play an unrenounceable role and is thus less important than
  $U$ itself.

  \section{Orbital degeneracy}\label{sec:zweiband}
  The properties of the metallic ferromagnets iron, cobalt, and nickel
  are determined by 3d electrons, implying a five-fold degeneracy.
  Therefore it has long been speculated that band degeneracy is an
  essential precondition for metallic ferromagnetism. Band degeneracy
  leads to additional on-site matrix elements of the Coulomb
  interaction describing intra-atomic interactions
  \begin{equation}\label{eqn:interband}
    V_{\mbox{\scriptsize interband}} = \sum_{i} \bigg[
    \sum_{\nu < \nu';\sigma\sigma'}  \! (V_0-\delta_{\sigma \sigma'}F_0)
    {n}_{i\nu\sigma}{n}_{i\nu'\sigma'}
    -F_0 \!  \sum_{\nu < \nu';\sigma \neq \sigma'}  \! 
    {c}^\dagger_{i \nu \sigma} {c}^{\phantom{\dagger}}_{i \nu \sigma'} 
    {c}^\dagger_{i \nu' \sigma'} {c}^{\phantom{\dagger}}_{i \nu'
      \sigma}\bigg] 
  \end{equation}
  as shown in Fig.~\ref{fig:hundillu} in the case of a two-fold
  degeneracy. In particular, they imply a density-density interaction
  $V_0$ and a (ferromagnetic) exchange interaction $F_0$ between
  electrons on different orbitals. These ``Hund's rule couplings'' are
  responsible for the ferromagnetic alignment of the spins on an
  isolated atom.  Slater~\cite{Slater36} and van
  Vleck~\cite{vanVleck53} suggested that this ``atomic magnetism'' may
  be transmitted from one atom to another by the kinetic energy,
  leading to coherent bulk order in the system.  The relevant
  Hamiltonian is then a sum of Hubbard models for each orbital,
  complemented by the purely local interband coupling terms in
  (\ref{eqn:interband}):
  \begin{figure}
    \epsfxsize=0.3\hsize\hspace{0.35\hsize}\epsfbox{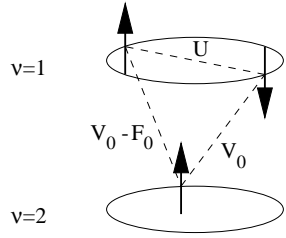}
    \caption{Illustration of the local interactions 
      between electrons in a
      two-band model.\label{fig:hundillu}}
  \end{figure}
  \begin{equation}\label{eqn:zweiband}
    H = \sum_{\nu}\bigg[ -t \sum_{\mbox{\scriptsize NN,} \sigma}  
    {c}^{\dagger}_{i \nu \sigma} {c}^{\phantom{\dagger}}_{j \nu \sigma} 
    + U \sum_{i}  {n}_{i\nu\uparrow}{n}_{i\nu\downarrow}
    \bigg] + V_{\mbox{\scriptsize interband}}.
  \end{equation}
  \noindent{}This model has received wide attention~\cite{2Band_a}, 
especially most
  recently~\cite{2Band_b}.  Away from quarter or half filling the model is
  particularly difficult to treat due to the high degeneracy in the
  atomic limit.  Quite generally ferromagnetism is found to be
  stabilized by Hund's rule coupling at intermediate to strong
  interactions. In this regime the DMFT, solved by QMC, once more
  provides a powerful method for the investigation of
  (\ref{eqn:zweiband})~\cite{Held98}.

  \begin{figure}
    \epsfxsize=0.6\hsize\hspace{0.2\hsize}\epsfbox{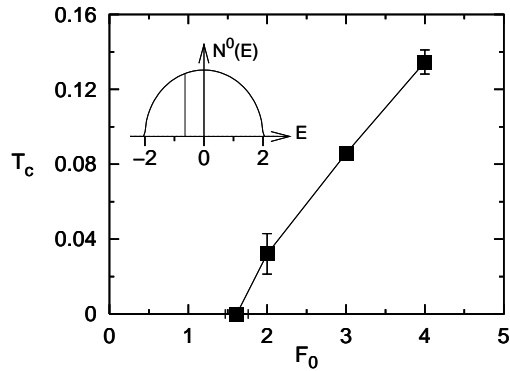}
    \caption{Curie temperature $T_c$ vs.~Hund's rule exchange coupling
      $F_0$ for a two-band Hubbard model with symmetric noninteracting
      DOS (see inset) as obtained within DMFT at $ U=9$, $V_0=5$, and
      $n=1.25$~\cite{Held98}.\label{fig:zweibandtc}}
  \end{figure}
\begin{figure}
    \leavevmode
    \epsfxsize=0.5\hsize\epsfbox{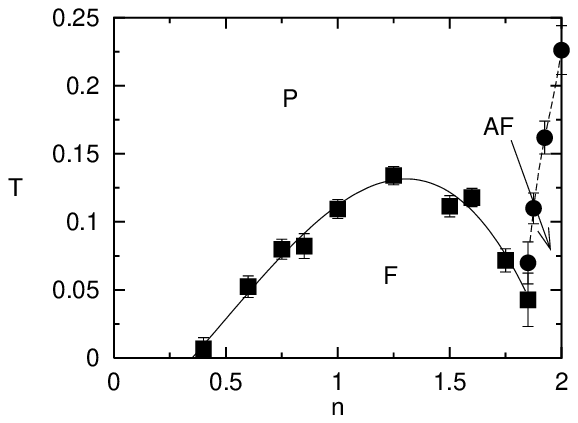}%
    \epsfxsize=0.5\hsize\epsfbox{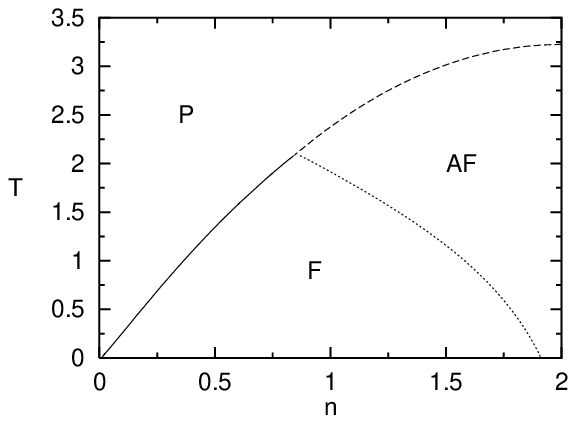}
\par\vspace{-1.5ex}\hspace{0.02\hsize}  \makebox[0.45\hsize]{\small (a)}%
    \hspace{0.08\hsize}%
    \makebox[0.4\hsize]{\small (b)}
    \caption{Magnetic phase diagram $T$ vs.~$n$ of a two-band Hubbard
      model for the same $U$ and $V_0$ as in
      Fig.~\ref{fig:zweibandtc}, and $F_0=4$: (a) DMFT, (b)
      Hartree-Fock~\cite{Held98}.\label{fig:zweibandphasdiag}}
  \end{figure}

  To identify the main mechanism responsible for ferromagnetism in the
  band-degenerate model and to distinguish it from that relevant for
  the one-band Hubbard model (namely, the strongly peaked DOS near the
  band edge) we here choose a featureless, symmetric Bethe-DOS as
  shown in the inset of Fig.~\ref{fig:zweibandtc}. For such a DOS no
  ferromagnetic instability was found in the one-band model up to the
  largest $U$ values within the DMFT~\cite{Wahle98} (see
  also~\cite{Obermeier97}).  In the following we restrict our
  discussion to a two-fold degeneracy. As can be seen from
  Fig.~\ref{fig:zweibandtc} no ferromagnetism occurs in the orbitally
  degenerate model even at $U=9$ if the Hund's rule exchange
  interaction $F_0$ is absent.  However, already a comparatively small
  value of $F_0$ is sufficient to make the ferromagnetic state
  favorable. The magnetic phase diagram $T$ vs.  $n$ is shown in
  Fig.~\ref{fig:zweibandphasdiag}a for the same interaction parameters
  as in Fig.~\ref{fig:zweibandtc} at $F_0 = 4$~\cite{Held98} (here we
  took into account the relation $U=V_0 + F_0$ which makes
  (\ref{eqn:zweiband}) form-invariant with respect to orbital
  rotations). Close to half filling ($n=2$) the antiferromagnetic
  state is found to be stable, while for lower filling ferromagnetism
  is stable in a broad range of densities. The maximum critical
  temperature is $T_c^{\mbox{\scriptsize max}} \sim 0.1$ which, for a
  band width of $4$~eV, corresponds to about $1000$~K. This result
  should be compared with the Hartree-Fock result
  (Fig.~\ref{fig:zweibandphasdiag}b) which is both qualitatively and
  quantitatively insufficient. In particular, Hartree-Fock theory does
  not describe the suppression of $T_c$ caused by the
  antiferromagnetic super-exchange near half filling.  Furthermore,
  the critical temperatures are by more than an order of magnitude too
  high, reflecting the absence of dynamical fluctuations in this
  approximation. Fig.~\ref{fig:zweibandtc} clearly shows that already
  moderately large Hund's rule couplings are able to mediate metallic
  ferromagnetism even in a system with an unspecific, symmetric DOS.
  It is interesting to see that the magnetic phase diagrams $T$ vs.
  $n$ for the one-band model (Fig.~\ref{fig:fcc3dim}a) and the
  band-degenerate model (Fig.~\ref{fig:zweibandphasdiag}a) look very
  similar, although the origin for the ferromagnetic phase is quite
  different.

  Here we did not discuss the possibility of orbital ordering where
  the electron densities alternate on the two orbitals on neighboring
  sites. In Ref.~\cite{Held98} it is found that orbital ordering sets
  in around quarter filling ($n=1$) when $F_0$ is decreased.

  \section{Discussion}\label{sec:concl}
  In this paper we discussed recent developments in our understanding
  of the origin of metallic ferromagnetism both in the one-band
  Hubbard model and the band-degenerate model. Analytical results for
  $d=1$ and, in particular, numerical results for $d=1,$ $2$, and
  $\infty$ were finally able to show convincingly that the one-band
  Hubbard model has a metallic ferromagnetic phase in a surprisingly large
  region of the on-site interaction $U$ and density $n$.  A
  stabilization of this phase at intermediate $U$ values requires a
  sufficiently large spectral weight near the band edge. Such a DOS is
  typical for frustrated lattices which optimize the kinetic energy of
  the polarized state and at the same time frustrate the parasitic
  antiferromagnetic ordering. By contrast, the origin of metallic
  ferromagnetism in the band-degenerate Hubbard model need not
  primarily be due to a DOS effect but is rather caused by (moderate)
  Hund's rule couplings. In this respect the origin of ferromagnetism
  in the orbitally degenerate model is more straightforward than in
  the one-band case. In the absence of orbital ordering the resulting
  magnetic phase diagrams are remarkably similar.
  
  The identification of a single main driving force for the
  stabilization of metallic ferromagnetism in the one-band and the
  band-degenerate model, respectively, helps to differentiate between
  different effects.  In real systems these effects will tend to
  conspire, as is evident, for example, in nickel where an fcc lattice
  leads to a strongly asymmetric DOS and the band degeneracy provides
  for Hund's rule couplings. The combination of these effects will be
  investigated in the future.

\end{document}